\documentclass{article}
\usepackage{spconf,amsmath,graphicx}

\title{Robust 3D Cell Segmentation: Extending the View of Cellpose}
\name{Dennis Eschweiler$^{1*}$\qquad Richard S. Smith$^{2}$\qquad Johannes Stegmaier$^{1*}$\thanks{*Correspondence should be addressed to \texttt{\{dennis.eschweiler},\texttt{ johannes.stegmaier\}@lfb.rwth-aachen.de}}}
\address{$^{1}$ Institute of Imaging and Computer Vision, RWTH Aachen University, Aachen, Germany\\
         $^{2}$ John Innes Centre, Norwich Research Park, Norwich, UK}
\begin{document}
%\ninept
%
\maketitle
\begin{abstract}
Increasing data set sizes of 3D microscopy imaging experiments demand for an automation of segmentation processes to be able to extract meaningful biomedical information.
Due to the shortage of annotated 3D image data that can be used for machine learning-based approaches, 3D segmentation approaches are required to be robust and to generalize well to unseen data.
The Cellpose approach proposed by Stringer \emph{et al.} \cite{stringer2020} proved to be such a generalist approach for cell instance segmentation tasks.
In this paper, we extend the Cellpose approach to improve segmentation accuracy on 3D image data and we further show how the formulation of the gradient maps can be simplified while still being robust and reaching similar segmentation accuracy. 
The code is publicly available and was integrated into two established open-source applications that allow using the 3D extension of Cellpose without any programming knowledge.
\end{abstract}
\section{Introduction} \label{sec:intro}
Current developments in fluorescence microscopy have allowed the generation of image data with progressively increasing resolution and often accordingly increasing data size \cite{economo2016,strnad2016}.
3D microscopy data sets can easily reach terabyte-scale, which requires a high degree of automation of analytical processes including image segmentation \cite{meijering2020}.
Machine learning-based segmentation approaches require annotated training data sets for reliable and robust outcomes, which often must be created manually and are rarely available.
Although some tools are available \cite{mcquin2018,spina18,dereuille2015,sommer2011}, this issue is even more severe for 3D microscopy image data, since manual annotation of 3D cellular structures is very time-consuming and often infeasible, which causes the acquisition of annotated 3D data sets to be highly expensive.

2D approaches have been applied in a slice-wise manner to overcome this issue \cite{stringer2020}, diminishing the need for expensive 3D annotations, but resulting in potential sources of errors and inaccuracies.
Segmentation errors are more likely to occur at slice transitions, leading to noisy segmentations, and omitting information from the third spatial dimension, \textit{i.e.}, a decreased field of view, leads to a potential loss of segmentation accuracy.
Conversely, if there are fully-annotated 3D data sets available, robust and generalist approaches are desired to make full use of those data sets \cite{stringer2020, isensee2021, schmidt2018}.

Stringer \emph{et al.} proposed the Cellpose algorithm \cite{stringer2020} and demonstrated that this approach serves as a reliable and generalist approach for cellular segmentation on a large variety of data sets.
However, this method was designed for 2D application.
Although a concept to obtain a 3D segmentation from a successive application of the 2D method in different spatial dimensions was proposed, the approach is still prone to the aforementioned sources of errors, missing on the full potential for 3D image data.
In this paper we propose (1) an extension of the Cellpose approach to fully exploit the available 3D information for improved segmentation smoothness and increased robustness.
Furthermore, we demonstrate (2) how the training objective can be simplified without loosing accuracy for segmentation of fluorescently labeled cell membranes and (3) we propose a concept for instance reconstruction allowing for stable runtimes.
(4) All of our code is publicly available and has been integrationed into the open source applications XPIWIT \cite{bartschat2016} and MorphoGraphX \cite{dereuille2015}.

\section{Method} \label{sec:method}
\begin{figure}[h]
    \centering
    \includegraphics[width=0.49\textwidth]{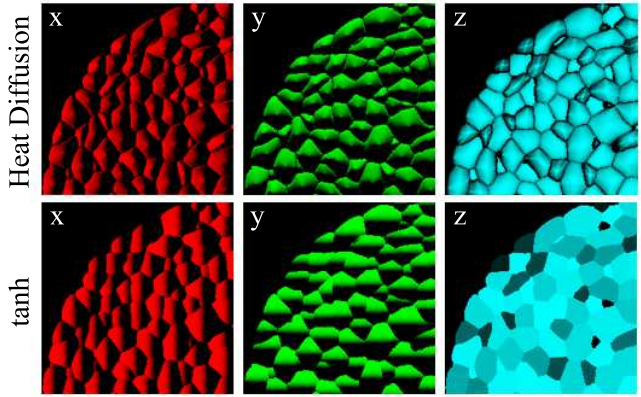}
    \caption{Gradient map representation of cell instances in $x$ (red), $y$ (green) and $z$ (blue) directions. The upper row shows heat diffusion simulations \cite{stringer2020}, while the lower row shows the simplified hyperbolic tangent distributions. Brightness indicates the gradient direction.}
    \label{fig:grad_maps}
\end{figure}

\begin{figure}[h]
    \centering
    \includegraphics[width=0.48\textwidth]{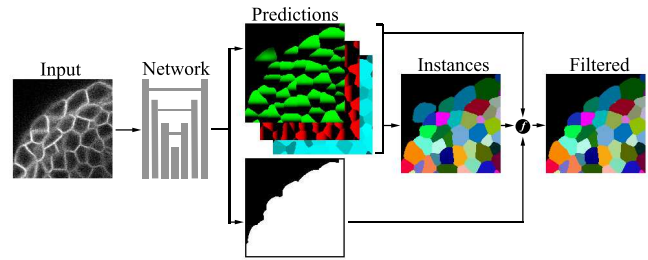}
    \caption{Representative visualization of our 3D processing pipeline, including the input patch \cite{wolny2020}, the neural network and the predicted outputs. Instances are reconstructed from the predicted gradient maps and filtered ($f$) by utilizing foreground predictions and flow errors.}
    \label{fig:pipeline}    
\end{figure}

The proposed method extends the Cellpose algorithm proposed by Stringer \emph{et al.} \cite{stringer2020}, which formulates the instance segmentation problem as a prediction of directional gradients pointing towards the center of each cell.
These gradients are derived from mathematically modelling a heat diffusion process, originating at the centroid of a cell and extending to the cell boundary (Fig.~\ref{fig:grad_maps}, upper row).
Gradients are divided into separate maps for each spatial direction, which can be predicted by a neural network alongside a foreground probability map to prevent background segmentations.
After prediction of the foreground and gradient maps using a U-Net architecture \cite{ronneberger2015}, each cell instance is reconstructed by tracing the multidimensional gradient maps to the respective simulated heat origin.
All voxels that end up in the same sink are ultimately assigned to the same cell instance.
For processing images directly in 3D and benefiting from the full 3D information, we propose multiple additions to the Cellpose approach and necessary changes to overcome memory limitations and prevent long run-times.
An overview of the entire processing pipeline is visualized in Fig. \ref{fig:pipeline}.

We propose to use a different objective for the generation of the gradient maps, since calculation of a heat diffusion in 3D (Fig.~\ref{fig:grad_maps}, top) is computationally complex.
Instead, we rely on a hyperbolic tangent spanning values in the range of $(-1,1)$ between cell boundaries in each spatial direction (Fig.~\ref{fig:grad_maps}, bottom), which can be interpreted as the relative directional distance to reach the cell center axis.
Note that the different formulation constitutes a trade-off between lower complexity and the ability to reliably segment overly non-convex cell shapes. 
Predictions are obtained by a 3D U-Net~\cite{cicek2016}, including all three spatial gradient maps for the $x$, $y$ and $z$ directions, respectively, and an additional 3D foreground map highlighting cellular regions.
Since the training objective is formulated by a hyperbolic tangent, the output activation of the U-Net is set to a hyperbolic tangent accordingly.
Processing images directly in the 3D space has the advantage of performing predictions only once for an entire 3D region, as opposed to a slice-wise 2D application, which has to be applied repetitively along each axis to obtain precise gradient maps for each of the three spatial directions.

Reconstruction of cell instances is done in an iterative manner, by moving each voxel within the foreground region along the predicted 3D gradient field $g$ by a step size $\delta_\textrm{recon}(x,y,z) = g(x,y,z) * s_\textrm{recon}$, given by the magnitude of the predicted gradient at each respective position $g(x,y,z)$ and a fixed integer scaling factor $s_\textrm{recon}$.
The number of iterations is defined by a fixed integer $N_\textrm{recon}$, which is adjusted to represent the number of steps necessary to certainly move a boundary voxel to the cell center.
Ultimately, each voxel ends up in the vicinity of the corresponding cell center, where they can be grouped and assigned a unique cell label.
Instead of utilizing clustering techniques to assign labels, we rely on mathematical morphology to identify connected voxel groups at each cell center.
As opposed to clustering, this allows for a constant and fast run-time, independent of the potentially large quantity of cells in 3D image data.
A morphological closing operation using a spherical structuring element with a radius $r_\textrm{closing}$ smaller than the estimated average cell radius is applied to determine those unique connected components.

Reconstructed cell instances are filtered by their size, requiring each instance to be within a predefined range of $(r_\textrm{min},r_\textrm{max})$ and by their overlap with the predicted foreground region with a required overlap ratio of at least $p_\textrm{overlap}$.
Similar to the post-processing used in \cite{stringer2020}, gradient maps are computed for each reconstructed instance and compared to the corresponding gradient maps predicted by the network. 
In an ideal case, both gradient maps are equivalent, but if their mean absolute error exceeds $err_\textrm{gradient}$ this instance is assumed to be falsely reconstructed and discarded from the final result.

\section{Experiments and Results} \label{sec:results}
\begin{figure*}[ht]
    \centering
    \includegraphics[width=0.49\textwidth]{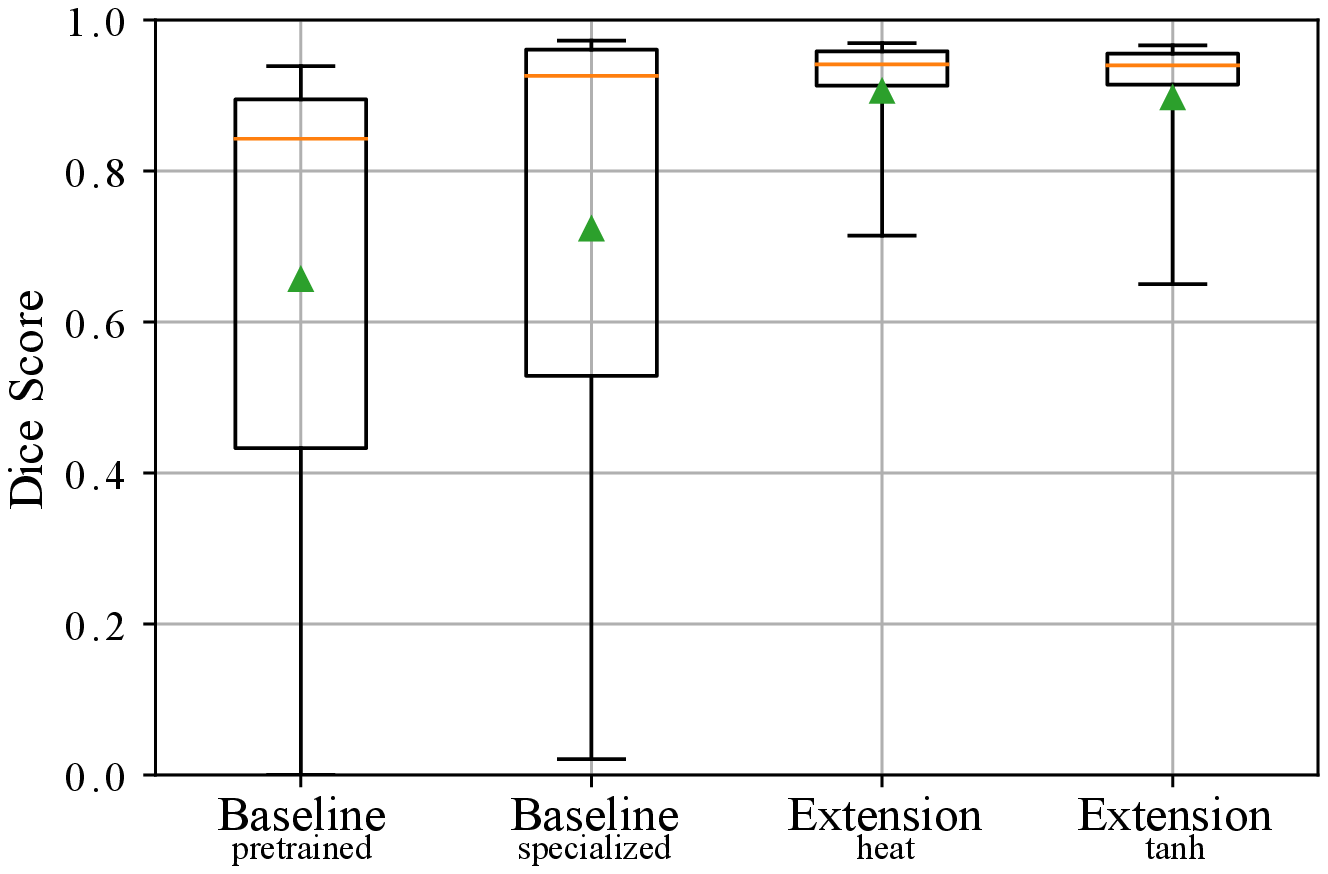}
    \includegraphics[width=0.49\textwidth]{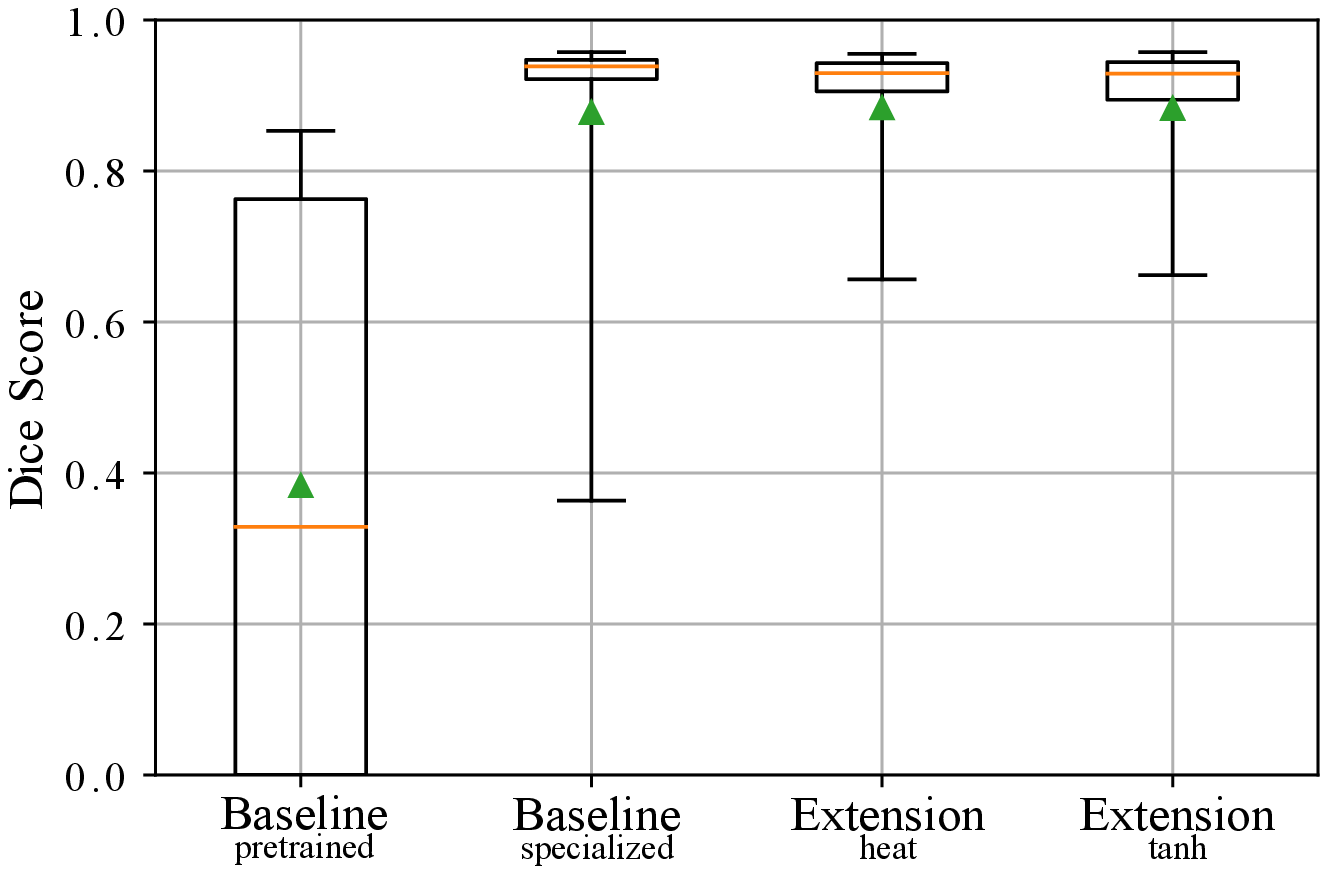}
    \caption{Dice scores computed from segmentation results obtained for the public Meristem data set \cite{willis2016} (left) and for the manually annotated 3D image stack (right). The experiments include pre-trained Cellpose provided by \cite{stringer2020} ($\text{Baseline}_\textrm{pretrained}$), a retrained version of the original Cellpose architecture ($\text{Baseline}_\textrm{specialized}$), our 3D extension with the original ($\text{Extension}_\textrm{heat}$) and the hyperbolic tangent gradient formulation ($\text{Extension}_\textrm{tanh}$). Whiskers represent the $5^\text{th}$ and $95^\text{th}$ percentile, the box spans from the $1^\text{st}$ to the $3^\text{rd}$ quartile, the orange line shows the median value and the green triangle indicates the mean value.}
    \label{fig:seg_scores}
\end{figure*}
To evaluate our method, we conduct different experiments comparing the 3D extension to the 2D baseline \cite{stringer2020} and assessing the generation of gradient maps.
For a comparison to various state-of-the-art segmentation approaches, we refer to the results reported in \cite{stringer2020}.
The data set we use for evaluation is publicly available \cite{willis2016} and includes $125$ 3D stacks of confocal microscopy image data showing fluorescently labeled cell membranes in the Meristem of the plant model organism \textit{A.~thaliana}.
We divided the data set into training and test sets, where plants 1,2,4 and 13 were used for training and plants 15 and 18 were used for testing.
The test data set comprises about 37,000 individual cells.
Furthermore, we show how well each of the approaches performs on unseen data to assess the generalizability to different microscopy settings.
Therefore, we use a manually annotated 3D confocal stack of \textit{A. thaliana}, comprising a total of 972 fully annotated cells.
Annotations were manually obtained using the \mbox{SEGMENT3D} online platform \cite{spina18}.
Experiments are structured as follows:
\vspace{-1em}\paragraph*{$\text{Baseline}_\textrm{pretained}$} As a baseline we use the publicly available Cellpose \textit{cyto}-model and apply it to the test data set using the 3D pipeline published in \cite{stringer2020}.
Since the model was trained on a large and highly varying data set, we do not perform any further training.
\vspace{-1em}\paragraph*{$\text{Baseline}_\textrm{specialized}$} The second experiment is designed to be a second baseline, as we use the original Cellpose approach \cite{stringer2020} and train it from scratch using the above-mentioned training data set from Willis \emph{et al.} \cite{willis2016} for 1000 epochs, which we find sufficient for convergence.
To reduce memory consumption and prevent high redundancies, we extract every fourth slice of each 3D stack of the training data set to construct the new 2D training data.
This experiment constitutes a specialized case of the baseline approach.
\vspace{-1em}\paragraph*{$\text{Extension}_\textrm{heat}$} Our proposed 3D extension is first trained with the original representation of the gradient maps, formulated as a heat diffusion process \cite{stringer2020}.
The 3D U-Net is trained on patches of size $128 \times 128 \times 64$ voxel for 1000 epochs using the Meristem training data set from Willis \emph{et al.} \cite{willis2016}.
In every epoch, one randomly located patch is extracted from each image stack of the training data set.
To obtain the final full-size image, a weighted tile merging strategy as proposed in \cite{de-bel19a} is used on the predicted foreground and gradient maps, before reconstructing individual instances.
Instance reconstruction parameters are empirically set to $s_\textrm{recon}=4$, $N_\textrm{recon}=100$ and $r_\textrm{closing}=3$.
Filtering parameters are set to $(r_\textrm{min},r_\textrm{max})=(5,100)$, $p_\textrm{overlap}=0.2$ and $err_\textrm{gradient}=0.8$.
\vspace{-1em}\paragraph*{$\text{Extension}_\textrm{tanh}$} For the fourth experiment, we change the representation of the gradient maps, formulated by hyperbolic tangent functions as described in Sec.~\ref{sec:method}.
Otherwise, the setup is identical to the setup of $\text{Extension}_\textrm{heat}$.\newline

Final mean Dice values for the public test data set \cite{willis2016} are $0.656$ for $\text{Baseline}_\textrm{pretrained}$, $0.723$ for $\text{Baseline}_\textrm{specialized}$ and $0.905$ and $0.897$ for $\text{Extension}_\textrm{heat}$ and $\text{Extension}_\textrm{tanh}$, respectively (Fig.~\ref{fig:seg_scores}, left).
Although  $\text{Baseline}_\textrm{specialized}$ benefits from specialized knowledge and shows improved segmentation scores, both baseline approaches result in poor instance segmentations in regions with low fluorescence intensity, leading to a large spread of obtained scores.
The full 3D extensions, however, are able to exploit the structural information from the third dimension to successfully outline poorly visible cell instances.
Furthermore, similar scores are obtained for both gradient formulations.
Results for the manually annotated data are shown as boxplots in Fig.~\ref{fig:seg_scores} (right) with mean Dice scores of $0.383$ for $\text{Baseline}_\textrm{pretrained}$, $0.877$ for $\text{Baseline}_\textrm{specialized}$ and scores of $0.883$ for both, $\text{Extension}_\textrm{heat}$ and $\text{Extension}_\textrm{tanh}$.
This confirms that the lack of structural 3D information leads to a loss in accuracy and that both gradient formulations perform similarly well.
\begin{figure*}[h!]
    \centering
    \includegraphics[width=1.0\textwidth]{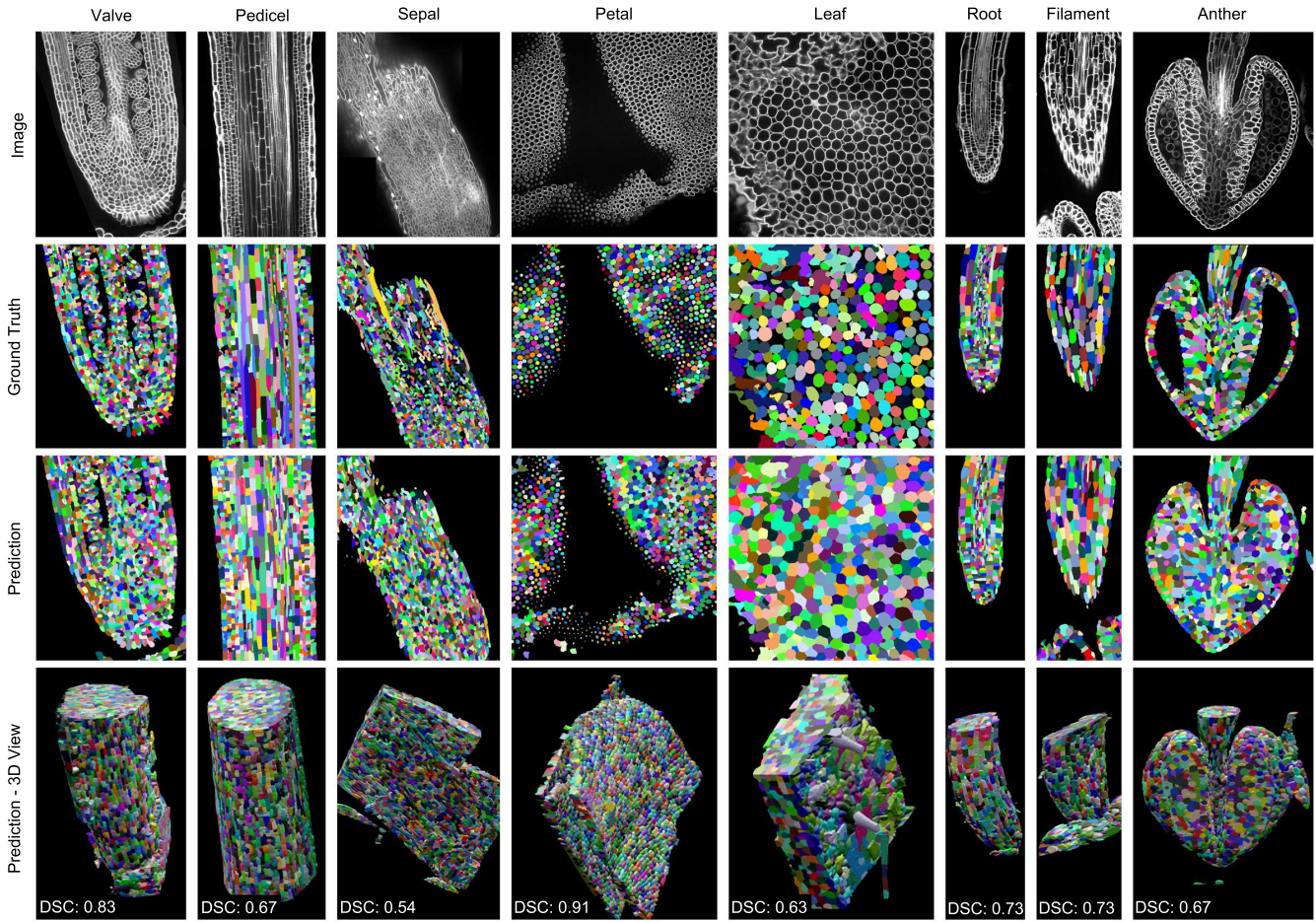}
    \caption{Slices of raw 3D image data (top), corresponding ground truth and predictions of the $\text{Extension}_\textrm{tanh}$ model (center) and 3D views of the predictions (bottom) including the obtained Dice scores (DSC) for different plant organs in \textit{A.~thaliana} \cite{wolny2020}.}
    \label{fig:atlas_seg}    
\end{figure*}

To further demonstrate the generalizability of the proposed approach, we generated results for the publicly available 3D image data from a variety of different plant organs in \textit{A.~thaliana} \cite{wolny2020}.
We used the model trained in $\text{Extension}_\textrm{tanh}$ and each image was scaled to roughly match the cell sizes of the training data set in each spatial direction.
Since the ground truth did not include segmentations of all cells being visible in the image data, we limited the computation of Dice scores to annotated cells.
Average Dice scores (DSC) obtained for instance segmentations of each image stack range from 0.54 to 0.91 and slices of raw image data, ground truth and instance predictions are shown in Fig.~\ref{fig:atlas_seg}.
This data set contains cellular shapes never seen during training, which causes overly elongated cells to be split up into different sections for some of the images.
Nevertheless, overall results demonstrate robustness and applicability to different cellular structures.

\section{Conclusion and Availability} \label{sec:conclusion}
In this work we demonstrated how the concept of Cellpose~\cite{stringer2020} can be extended to increase segmentation accuracy for 3D image data.
The utilization of the full 3D information and the prediction of 3D gradient maps, help to improve segmentation of cells in regions of poor image quality and low intensity signals.
Our alternate formulation of the proposed gradient maps leads to a comparable accuracy of segmentation results, while offering a lower complexity with respect to training data preparation and instance reconstruction.
The morphology-based approach proposed for the instance reconstruction enables the application to 3D microscopy image data independent from the quantity of captured cells.
Results obtained on completely different data sets never seen during training support the claim that this approach is generalist and robust.
Code, training and application pipelines are publicly available at https://github.com/stegmaierj/Cellpose3D.
Furthermore, we integrated the approach into the existing open source applications XPIWIT \cite{bartschat2016} and MorphoGraphX \cite{dereuille2015} to make it accessible to a broad range of community members.

%\section{COMPLIANCE WITH ETHICAL STANDARDS}
%The work deals with publicly available image data for which no ethical approval was required.

\section{ACKNOWLEDGEMENTS}
This work was funded by the German Research Foundation DFG with the grants STE2802/2-1 (DE) and an Institute Strategic Programme Grant from the BBSRC to the John Innes Centre (BB/P013511/1).

\vfill\pagebreak

\bibliographystyle{IEEEbib}
\bibliography{main}

\end{document}